\documentclass{article}
\usepackage[utf8]{inputenc}
\usepackage[hyperref]{acl2021}
\usepackage{times}
\usepackage{latexsym}

\usepackage{dialogue}
\usepackage{fancyvrb}
\usepackage{graphicx}
\usepackage{wrapfig}
\usepackage{csvsimple}
\usepackage{tabulary}
\usepackage{amssymb}%
\usepackage{pifont}%

\usepackage{microtype}

\usepackage{newfloat,caption,float}

\DeclareFloatingEnvironment[
  fileext = loe,
  listname = Examples,
  name = Example,
  placement = H,
  within = none,
  ]{example}
\usepackage{listings}
\usepackage{xcolor}

\colorlet{punct}{red!60!black}
\definecolor{background}{HTML}{EEEEEE}
\definecolor{delim}{RGB}{20,105,176}
\colorlet{numb}{magenta!60!black}
\lstdefinelanguage{json}{
    basicstyle=\normalfont\ttfamily,
    numbers=left,
    numberstyle=\scriptsize,
    stepnumber=1,
    numbersep=8pt,
    showstringspaces=false,
    breaklines=true,
    frame=lines,
    backgroundcolor=\color{background},
    literate=
     *{0}{{{\color{numb}0}}}{1}
      {1}{{{\color{numb}1}}}{1}
      {2}{{{\color{numb}2}}}{1}
      {3}{{{\color{numb}3}}}{1}
      {4}{{{\color{numb}4}}}{1}
      {5}{{{\color{numb}5}}}{1}
      {6}{{{\color{numb}6}}}{1}
      {7}{{{\color{numb}7}}}{1}
      {8}{{{\color{numb}8}}}{1}
      {9}{{{\color{numb}9}}}{1}
      {:}{{{\color{punct}{:}}}}{1}
      {,}{{{\color{punct}{,}}}}{1}
      {\{}{{{\color{delim}{\{}}}}{1}
      {\}}{{{\color{delim}{\}}}}}{1}
      {[}{{{\color{delim}{[}}}}{1}
      {]}{{{\color{delim}{]}}}}{1},
}

\aclfinalcopy %

\title{EMISSOR: A platform for capturing multimodal interactions as Episodic Memories and Interpretations with Situated Scenario-based Ontological References}

\author{Selene B\'{a}ez Santamar\'{i}a, Thomas Baier, Taewoon Kim, Lea Krause, Jaap Kruijt, Piek Vossen\\
Computational Linguistics and Text Mining Lab (CLTL)\\
Vrije Universiteit Amsterdam, Netherlands\\
\texttt{s.baezsantamaria,t.baier,t.kim,l.krause,j.m.kruijt,p.t.j.m.vossen@vu.nl}\\
}
\date{April 2021}

\begin{document}

\maketitle

\begin{abstract}
We present EMISSOR: a platform to capture multimodal interactions as recordings of episodic experiences with explicit referential interpretations that also yield an episodic Knowledge Graph (eKG). The platform stores streams of multiple modalities as parallel signals. Each signal is segmented and annotated independently with interpretation. Annotations are eventually mapped to explicit identities and relations in the eKG. As we ground signal segments from different modalities to the same instance representations, we also ground different modalities across each other. Unique to our eKG is that it accepts different interpretations across modalities, sources and experiences and supports reasoning over conflicting information and uncertainties that may result from multimodal experiences. EMISSOR can record and annotate experiments in virtual and real-world, combine data, evaluate system behavior and their performance for preset goals but also model the accumulation of knowledge and interpretations in the Knowledge Graph as a result of these episodic experiences.
\end{abstract}

\section{Introduction}

Multimodal interaction in real-world settings (physical or virtual) using (simulated) sensors between humans and robots as agents is a complex process. Furthermore, it typically evolves over time and within a shared (physical) space, being bound in both time and place, yet remaining continuously dynamic. The fact that certain contextual factors are not physically present, such as past episodic encounters, background knowledge and (hidden) intentions, adds to this complexity. Agents designed to behave intentionally need to handle this complexity and form teams with people to collaborate and achieve shared goals.

Within the Hybrid Intelligence framework,\footnote{\url{www.hybrid-intelligence-centre.nl}} we are specifically interested in such collaborative settings and and focus on analysing what causes such systems to succeed or fail within these. Collaboration requires shared grounding and partially shared understanding of situations, communications, and references across modalities. As humans and agents may have different beliefs and perceptions of these situations, we argued in previous work \citep{vossen2018leolani, vossen_leolani_2019} that agents need a theory-of-mind (ToM) \citep{premack1978does, leslie1987pretense} to handle conflicts, miscommunication and errors in referential grounding and interpretations.\footnote{\url{makerobotstalk.nl}}

Although there are many initiatives for representing multimodal interactions, referential grounding is hardly handled in its full complexity. Most approaches to multimodal interaction data either \textit{label} media files such as video, images, or audio with annotations or simply present situated agent interactions in the form of dialogues or their actions without labeling. In these approaches, annotations may be seen as interpretations of direct ``behavioral" responses (utterances or actions) to the preceding signals. However, they lack a formalization of these interpretations of experiences into an explicit model that supports transparent reasoning. Such multimodal data sets can be seen as episodic experiences but not yet as knowledge-aware episodic memories that reflect these experiences' cumulative results. The latter requires interpretations of different multimodal signals to be combined in an explicit knowledge structure according to an ontological model that reflects our conceptualization of the world. In addition, this model needs to be able to handle alternative interpretations, uncertainties and conflicts as the interpretations are not always correct or consistent.

We propose a generic model that can capture multimodal interactions as recordings of episodic experiences with explicit referential interpretations that also yield an episodic Knowledge Graph (KG) as a ToM . \textbf{EMISSOR} stands for \textbf{E}pisodic \textbf{M}emories and \textbf{I}nterpretations with \textbf{S}ituated \textbf{S}cenario-based \textbf{O}ntological \textbf{R}eferences. The platform stores streams of multiple modalities as parallel signals. Each signal can be segmented  and annotated independently with interpretation, providing robustness and simplicity. Stored signals can thus represent natural conversations in situated contexts in which (visual) actions and (verbal) utterances can be responses to each other but can also happen independently.  EMISSOR can represent any (multimodal) interaction that takes place either in virtual or real-world settings, involving any virtual or real-world agent.

Annotated signals do not necessarily stand on their own, but may be mapped to explicit identities, relations, and properties in an episodic Knowledge Graph (eKG) for capturing time-series of instances of situations. These mappings ground signal segments to formal instance representations and ground different modalities across each other. Time-bound experiences are thus captured as episodic experiences, i.e. as an explicit cumulative interpretation of streams of signals. The eKG models knowledge and interpretation shifts over time and supports reasoning over the interpretation. By keeping track of the provenance of signals and interpretations, our model reflects alternative ToM interpretations from different sources, modalities and experiences.

In the current paper, we set out the basic design and structure of our representation and our motivation. We first discuss in Section~\ref{sec:related-work} representations of multimodal interaction proposed in various paradigms such as virtual games, agent interactions, multimodal dialogue systems. In Section~\ref{sec:design}, we describe the desiderata for our proposal for representing interactions, which combines aspects from the different approaches in the related work but adds a KG as the host of such transparent episodic memories for situated experiences. We elaborate on the different data layers and relations in our proposal. We demonstrate how different types of data sets can be converted and represented in EMISSOR in Section~\ref{sec:examples}. In Section~\ref{sec:annotation}, we show how such multimodal data can be aligned and annotated such that segments get grounded to identities in a Knowledge Graph using an annotation tool. Future research and conclusions are presented in the final  Section~\ref{sec:conclusion}.

\section{Related work}
\label{sec:related-work}
Interaction data can take many forms. Not only can it come in different modalities (visual, audio, text, action), but we can also have many different types of interactions, e.g., search, question-answering, command-action sequences, (task-based) dialogues, navigation tasks, games, graphical interfaces, plain video, and audio recordings. It is impossible to provide a comprehensive overview of data representations in each separate modality and interaction type in this paper.

For our research on social communicating robots, we are interested in representations of multimodal interactions with \textit{referential grounding} across modalities and the representations of these modalities as such. Therefore, we discuss mainly works on modalities aligned in time series representing interactions. This excludes data in single modalities, such as plain text corpora with dialogues (spoken or text) and image or video data without dialogues. It also excludes static data that does not represent temporal sequences of data. For example, visual data labeled with textual descriptions and textual data augmented with visual scenes do not necessarily represent interactions. In interactions, modalities partially complement each other, such as speech responding to speech or to scenes and actions following speech. Such sequences often react to and complement each other and reflect some degree of causality and coherence, but not entirely. Augmented modalities, on the other hand, mainly represent paired data where one modality describes or illustrates the other. Collections of augmented data, e.g., labeled Flickr images, do not exhibit coherence across data points and do not reflect causal interaction. Nonetheless, single modality data and non-interactive data can still aid in processing multimodal interaction. Models and classifiers trained on such data can automatically annotate scenes in a time series of multimodal interaction data. An interesting research question would be whether static or single modality annotations also model collaborative interactions without considering temporal, causal, and coherence relations across data.

A recent classification and overview of interactive dialogue datasets is given in \cite{serban2018survey}. \citet{serban2018survey} differentiate dialogue systems on the types of interaction (goal(s), non-goal chit-chat, topical); their modalities (written, spoken, video); the participants (Human, Agent); being constrained, spontaneous, scripted, fictional; being goal-oriented, domain-specific or open. Most of the data described is, however, non-situated or not situation-grounded. There is hardly any reference to and interaction with physical contexts. Many of these datasets and tasks have been developed and presented in SigDial\footnote{\url{www.sigdial.org}}, the ACL Special Interest Group, for research on dialogues structures and models. A more applied perspective is taken by the Dialogue System Technology Challenge (DSTC\footnote{\url{dstc9.dstc.community}}). DSTC provides a platform for researchers and industry to develop and evaluate agent-interaction systems. Older datasets mostly contain conversational data for chatbots. A more recent challenge, Audio Visual Scene-Aware Dialog (AVSD, \cite{alamri2019audiovisual}), though contains short video clips with audio, descriptive captions, and a dialogue history. Participating systems need to answer a follow-up question or ground an answer to an image from the video or an audio fragment. This challenge represents situated-references and grounding across modalities, but the conversation is very descriptive: the conversations describe situations rather than being embedded in them. Below, we briefly describe some well-known datasets and challenges that represent various types of interaction data.

ParlAI\footnote{\url{github.com/facebookresearch/ParlAI}} released more than a hundred conversational datasets covering a wide range of topics, but most are single modality chat \cite{miller2017parlai}. SIMMC\footnote{\url{github.com/facebookresearch/simmc}} is Facebook's sequel to ParlAI with Situated and Interactive Multi-Modal Conversation \cite{crook2019simmc}. It consists of task-oriented dialogues in multimodal contexts represented by collections of images. The data contains referential relations between the dialogues and situations, but the current challenge is restricted to the e-commerce contexts of buying furniture and fashion items. There is no grounding to complex situations but intentions and goals are explicitly represented. Facebook-research also launched various other related tasks, among which RECCON\footnote{\url{github.com/declare-lab/RECCON}}: Recognizing Emotion Cause in Conversations \cite{poria2020recognizing}, and MINIRTS:  Hierarchical Decision Making by Generating and Following Natural Language Instructions in Real-time strategy game environments\footnote{\url{github.com/facebookresearch/minirts}} \cite{hu2019hierarchical}. The former grounds dialogues to emotions and their causes but not to visual or audio data. The latter grounds language to a closed virtual world by references to objects, agents, and actions. The dialogues are limited to commands and instructions to operate the game.

Google developed Schema-Guided-Dialogue (SGD\footnote{\url{github.com/google-research-datasets/dstc8-schema-guided-dialogue\#dialogue-representation}}) as a framework for task-oriented conversational agents \cite{rastogi2020towards}. In addition to e-commerce services, the tasks involve intent prediction, slot filling, dialogue state tracking, policy imitation learning, language generation, and user simulation learning. The goals are defined, but there is no multimodal situational grounding.

An older comprehensive robot platform is provided by openEASE\footnote{\url{www.open-ease.org}}, which is a web-based knowledge service providing robot and human activity data constituting episodic memories \cite{beetz2015open}. It produces semantically annotated data of manipulation actions, including the agent's environment, the objects it manipulates, the task it performs, and the behavior it generates. %
EASE uses so-called NEEMS (Narrative Enabled Episodic Memories) as episodic memories. NEEMS consist of a video recording by the agent of the ongoing activity. These videos are enriched with a story about the actions, motion, their purposes, effects, and the agent’s sensor  information during the activity.%
EASE is not data-centric but a service platform that uses a knowledge database as a back-end. The database can be explored through prolog queries. %
The focus of openEASE is on physical interactions and not on conversations with complex referential relations between expressions and situations.

Microsoft created a Platform for Situated Intelligence, PSI\footnote{\url{github.com/Microsoft/psi}} \cite{bohus2017rapid}. PSI offers multimodal data visualization and annotation tools, as well as processing components for various sensors, processing technologies, and platforms for multimodal interaction. PSI is a complete solution for modeling multimodal situations and interactions within. It comes closest to a comprehensive solution. However, PSI is mainly a software integration platform through which developers can share modules using a streaming architecture for signal annotation. The interactions are not stored in a shared representation, and the platform cannot be used for sharing experimental data independent of the platform developed in .NET.

Action Learning From Realistic Environments and Directives (ALFRED\footnote{\url{askforalfred.com}}) is a recent benchmark for learning a mapping from natural language instructions and egocentric vision to sequences of actions for household tasks \cite{ALFRED20}. ALFRED releases challenges and leader boards for trajectory tasks in which a human instructs an agent through natural language commands to carry out specific household tasks in a virtual world. The task is grounded in situations and combines video, audio, and text for clear goals. There is no natural dialogue that is independent of the task.

The previous datasets involve agents and, in some cases, the virtual world. Video recordings of interacting people can be seen as another source of data. MELD\footnote{\url{affective-meld.github.io/}} and COSMIC\footnote{\url{github.com/declare-lab/conv-emotion}} represent the Friends sitcom through videos, time-stamped dialogues and emotion annotations \cite{poria2019meld, ghosal2020cosmic}. Two seasons from the Friends dialogues were also annotated with person references and identities by \citep{choi-chen-2018-semeval} for the SemEval2018-task4 on and for Q\&A on open dialogues \cite{yang-choi-2019-friendsqa}.
IEMOCAP\footnote{\url{sail.usc.edu/iemocap/}} created detailed multimodal recordings of scripted human-human conversations with annotations of participants, gaze, gestures, etc. \cite{busso2008iemocap}. Both datasets do not provide any further situated-references, and there is no specific goal set for the conversation. Other similar smaller datasets with audio-visual emotion expression are RAVDESS~\cite{livingstone2018ryerson}, TESS~\cite{dupuis2010toronto} and SAVEE~\cite{HaqJackson_MachineAudition10}.

\subsection{Annotation schemes}
Interaction data come in different formats and follow different schemes. DiaML \cite{bunt2012using} is a modeling language for the annotation of dialogues as a discourse. However, it does not tackle the grounding problem. It does not make use of identifiers that represent identities of entities, contexts, and situations independently of their mentions as it targets single modality data.

VOXML \cite{pustejovsky2016voxml}  is a formal modeling language for capturing spatial semantics of object entities in 3D simulations. VOXML tackles grounding but does not model dialogue interaction nor the sequential alignment of cross-modality segments. It is a model for defining the semantics of linguistic expressions through physical world simulations, and it does not model this world per se independently of these expressions. Furthermore, it is a formal symbolic representation that relies on a fully descriptive relation between language expressions and situation modeling. %

The Simple Event Model or SEM is a Resource Description Framework (RDF) model for capturing situations following semantic web principles \cite{van2011design}. SEM represents situations as event instances through URIs, with actors, places, and temporal relations to OWL-Time objects\footnote{\url{www.w3.org/TR/owl-time/}} either defined as time points or as periods. Situations can be related as sequences in time series through OWL-Time grounding, as well as through explicit temporal and causal relations between events. SEM can be used to construct event-centric KGs rather than entity-centric KGs. Event-centric KGs are well-suited for representing temporal properties of situations and entities within these. Furthermore, they are not limited to the predefined properties of entity-centric graphs but exploit abstract event-participant relations that can be further modeled in additional ontologies \cite{segers-etal:2018}.

The Grounded Representation and Source Perspective (GRaSP) model \cite{fokkens2017grasp}, augments SEM with \textit{grasp:denotes} relations between linguistic expressions (so-called mentions) and their referential identities. Through GRaSP, any segment in a signal (verbal, audio, or video) can be mapped to an instance in a SEM model, as such providing a flexible framework for referential grounding. Although SEM and GRaSP can be used for any modality, they have mostly been used for representing events in text \cite{vossen2016newsreader}. In \cite{vossen2018leolani, vossen_leolani_2019}, we have shown that GRaSP can also be used for modeling multimodal situations with unaligned signals to model a theory-of-mind or ToM in which different modalities and different sources can generate alternative interpretations that can co-exist in the robot's eKG. Similar to \cite{kondratyuk2017towards}, our robot eKG reflects the episodic accumulation of knowledge through interaction over time. Our model differs from theirs in that our model allows for alternative facts and properties.

DIAML, VOXML and GRaSP are complex XML representations. Most of the DSTC datasets, however, follow a more basic schema in JSON format. Data elements represent sequences, possibly including time stamps and pointers to separate media files, possibly including bounding box coordinates, scene interpretations, participants, utterances, and goals to achieve. Google's Schema-Guided-Dialogue, Amazon's Alexa Topical-Chat and Facebook's SIMMC provide comparable JSON formats for capturing simple situations, goals, participants and dialogues with communication.

\subsection{Our contribution}
Although we can use many aspects of the previously discussed data and representation models, none of these completely provide what we need for modeling streaming data in different modalities representing parallel sequences of signals within physical or virtual world contexts while allowing for alternative interpretations.
Most of the described data and models do not entirely represent the contextual situation in which conversations are embedded. Furthermore, they lack the means to represent the cumulative result of interpreting streams of multimodal signals over time.

In our representation, we combine the best of two worlds. On the one hand, we use light-weight JSON-LD\footnote{\url{json-ld.org/}} for data in different modalities, their segmentation, alignment, and annotations, which provides us with the flexibility to easily represent any data streams; on the other hand, we use RDF to represent interpretations of such multimodal situations and the referential relations to explicit identities following GRaSP, which allows us to reason over the data in a robust way. Likewise, our framework can model the interpretation from raw signals to interpreted segments up to the situation-centric aggregation of triples over time as an episodic memory.

Our approach to connecting multimodal situated interactions to an explicit Knowledge Graph comes close to openEASE \cite{beetz2015open}, except that we focus on complex \textit{reference} relations between conversations and situations rather than on robot \textit{actions} only. Furthermore, we follow a data-centric approach in which an open representation of the interaction forms the basis for sharing data, tools, and solutions, whereas interactions in openEASE can only be accessed through queries. %
In contrast to the episodic triples from conversation generated by \cite{kondratyuk2017towards} and the episodic NEEMS of openEASE, our eKG incorporates a ToM model based on SEM and GRaSP which supports reasoning over conflicting information, knowledge gaps and uncertainties across modalities and sources.

Nevertheless, EMISSOR is not restricted to a specific annotation scheme nor a specific formal model of situations. The use of JSON-LD allows seamless integration with any eKG, i.e. from raw signal, to annotation, to explicit symbolic representation. Our EMISSOR platform supports converting different datasets to a shared model that imposes alignments of spatial and temporal signals within a referential grounding framework and an episodic memory with understanding. The framework combines modalities, generates segmentation for each, and creates referential grounding and corresponding triple representations that capture identities and relations. We furthermore provide an annotation tool to create gold annotations for referential grounding, both from dynamic streams and manually designed scenarios using controlled static data.

\section{EMISSOR: design and specification}
\label{sec:design}

In this section, we describe the different data layers of our model and their interrelations. We summarise the design desiderata for our model as follows:

\begin{enumerate}
\itemsep0em
    \item Support parallel unaligned streams of multimodal signals
    \item Detect sequences of segments within signals grounded in time and space
    \item Allow segment alignment, overlap, and disjointedness across modalities
    \item Model situated-references in segments to unique identities in a Knowledge Graph
    \item Model causal coherent relations between subsets of segments across modalities
    \item Model physical and virtual real-world contexts
    \item Model streams of signals as a transparent cumulative symbolic interpretation of experiences
    \item Provide an episodic memory of situated references and interpretations in a knowledge graph that supports reasoning
\end{enumerate}

To meet these requirements, we use GRaSP as a referential framework to connect segments in signals to identities in an eKG. These identities are individual people, objects, and places, or relations and properties of these. In the former case, identities are represented as instances through their URIs, possibly with names as labels and instances of a particular type. In the latter case, constellations of instances are interpreted as relations or properties represented by RDF triples. In addition, media streams are stored as separate data files for each modality. Separate JSON-LD files are provided for each modality that 1) divide this modality signal into segments grounded in time and space and 2) annotate these segments as representing identities and relations registered in the eKG. A stream of media interpretations emitting triples over time then results in the cumulative growth of an eKG representing an episodic memory. With the use of JSON-LD elements, signal metadata can be mapped to our referential framework and included in the eKG.

\paragraph{Motivating example}
Let us consider a simple example (\ref{ex1}). A face detector module detects a human face in a video frame at time \textit{t1} which results in a box segment. Face recognition cannot recognize this person, so it is identified as a new instance of the type \textit{PERSON}. Next, at \textit{t2}, another face is detected as a box segment. This face is identified as a known person in the eKG: a URI with the name `Carl'. At \textit{t3} a speaker is detected whose voice is mapped to the same identity of ``Carl". The speaker says ``Do you know my daughter Carla?". A text understanding module processes the text representation of the audio at \textit{t3}. Words such as ``I" and ``me" are linked to the same URI as the speaker, while ``Carla" will be mapped to another identity (not yet visually grounded) and related to ``Carl" as his daughter. Through contextual reasoning, the model may ground the first unidentified face perception to this newly created identity of ``Carla" in hindsight.

\begin{example}
\begin{verbatim}
video, t1 -> segment -> [PERSON]
video, t2 -> segment -> URI(:Carl)
audio, t3 -> segment -> URI(:Carl)
text,  t3 -> segment -> URI(:Carl)
text,  t3 -> segment -> URI(:Carla)
text,  t3 -> segment -> triple
           (:Carl :daughter :Carla)
reinterpretation:
video, t1 -> segment -> URI(:Carla)
\end{verbatim}
\caption{Carl - Carla scenario}
\label{ex1}
\end{example}

Let us consider an alternative variation on this scenario in which the detected speaker at \textit{t3} is mapped to ``Alice" rather than ``Carl". Alice says, ``That is Carl and his daughter Carla". The deictic references to ``Carl" and ``Carla" and the pronominal reference ``his" can only be resolved by combining the earlier perceptions with the semantics of this utterance. On the other hand, in both scenarios, ``being a daughter" is knowledge that cannot be concluded from visual and audio signals and is solely conveyed by interpreting the semantics of the utterance. This demonstrates that neither audio-visual nor textual segments contain all the information needed to come to the correct interpretation: language utterances due to their referential nature rather than being descriptive,  audio-visual segments due to their limitation to signal social, conceptual, and cultural framing.

\subsection{Model description}

\begin{figure*}
  \centering
  \includegraphics[width=\linewidth]{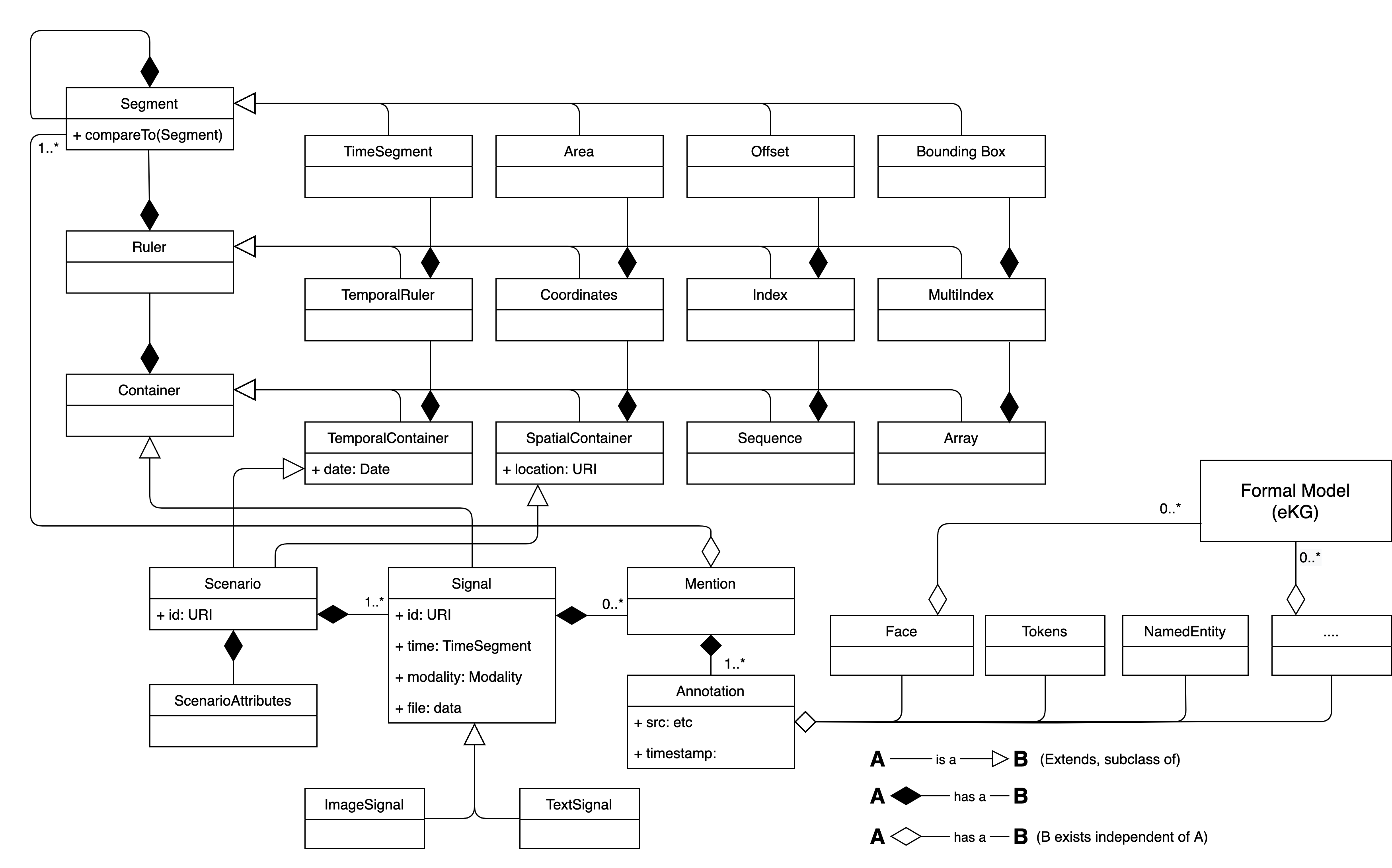}
  \caption{Entity-relationship overview of data elements and relations within EMISSOR. The right-side box functions as a placeholder for any formal model of situations that can be linked to the annotations of segments in the multimodal data streams. We assume here that identities in these models can be defined according to any set of ontologies to reason over the interpretations.}
  \label{fig:overview}
\end{figure*}

Figure \ref{fig:overview} shows an overview of our representation's different data layers in terms of an entity-relationship model. For grounding data, we define different layers for segments, rulers, and containers. Segments can have complex causal coherence relations across modalities. Since the segmentation of these modalities is done separately, we need temporal and spatial containers to order and connect segments across modalities, following the principles of TimeML \cite{pustejovsky2011increasing} and VOXML. The temporal and spatial containers form the basis for constituting potential causal relations (forward and backward and across distances). Therefore, each container will consist of a ruler that defines the granularity of segments across modalities (e.g., a sequence or region). The ruler positions the segments relative to each other and makes them conditional for defining relations and for predictive models.

A scenario (bottom left) is an instance of a context in a specific time and space and acts as a container for parallel streams of multimodal signals, which are divided into segments. Scenarios can have specific attributes to qualify them, including names, overall scenario type and location, the purpose or intention of (specific) participants. When segments get annotated, mentions (linguistic) or perceptions (video or audio) are created, pointing to one or more segments and an annotation value. Annotations can be added freely and there are no restrictions on the values for annotations in the JSON structure. Figure~\ref{fig:overview}, shows a few examples of typical annotation values such as \textit{Face} for boxes in images, \textit{Tokens} in texts or \textit{NamedEntity} for Named Entity expressions. However, EMISSOR additionally uses JSON-LD to also support the direct linking of interpretations to the eKG in which people, objects and situations are modeled through explicit URIs. We therefore allow explicit URIs as annotation values to ground the segments to these identities. In that case, a box segment, a name or pronoun in the text is annotated with the unique URI of a person rather than a conceptual label as a value.

In Figure~\ref{fig:overview} due to limits of space, we only show a place holder for the eKG as a formal model of situations at the right side. This place holder stands for any ontological model and its population with instances. In practice, we use a wide range of ontologies to model situations as populations of instances, as described in \cite{context19-leolani}. Note that EMISSOR allows for both types of annotations next to each other, e.g. a segment can be annotated as a human face without a specific identity but the same segment can have an additional annotation with the URI from the eKG.

\paragraph{Scenario structure}
We consider an interaction as a scenario. Scenarios are organized in folders on disk. Within a scenario folder, we store the source data as separate files in a modality subfolder, e.g. text, video, image, audio. Furthermore, one JSON-LD file per modality defines the metadata and the segments present for each signal in the modality. %
In addition to these segments, there may be lists of mentions or perceptions as JSON-LD elements. Each mention or perception specifies a range of segments (at least one) and the interpretations as annotations representing the instances and concepts in Figure \ref{fig:overview}. Next to the modality JSON-LD file, a specific folder contains the RDF triples extracted from the annotated signals. For example, an utterance in a conversation may mention somebody's age, which yields an RDF triple with the person's URI as the subject, the \textit{has-age} property, and the actual age as a value.

Finally, there is a separate JSON file with metadata on the complete scenario. This scenario JSON defines the temporal and spatial ruler within which the scenario is located (date, begin and end time, geo-location, place-name), the interacting participants (e.g., the agent and the human speaker(s)), and any other people and objects that participate in the scene. The specification of participants and props can be based on the instances from the eKG. This scenario JSON file has the same name as the folder name of the scenario. Example~\ref{ex2} shows the file structure for a scenario.

\begin{figure*}
  \centering
  \includegraphics[width=0.95\textwidth]{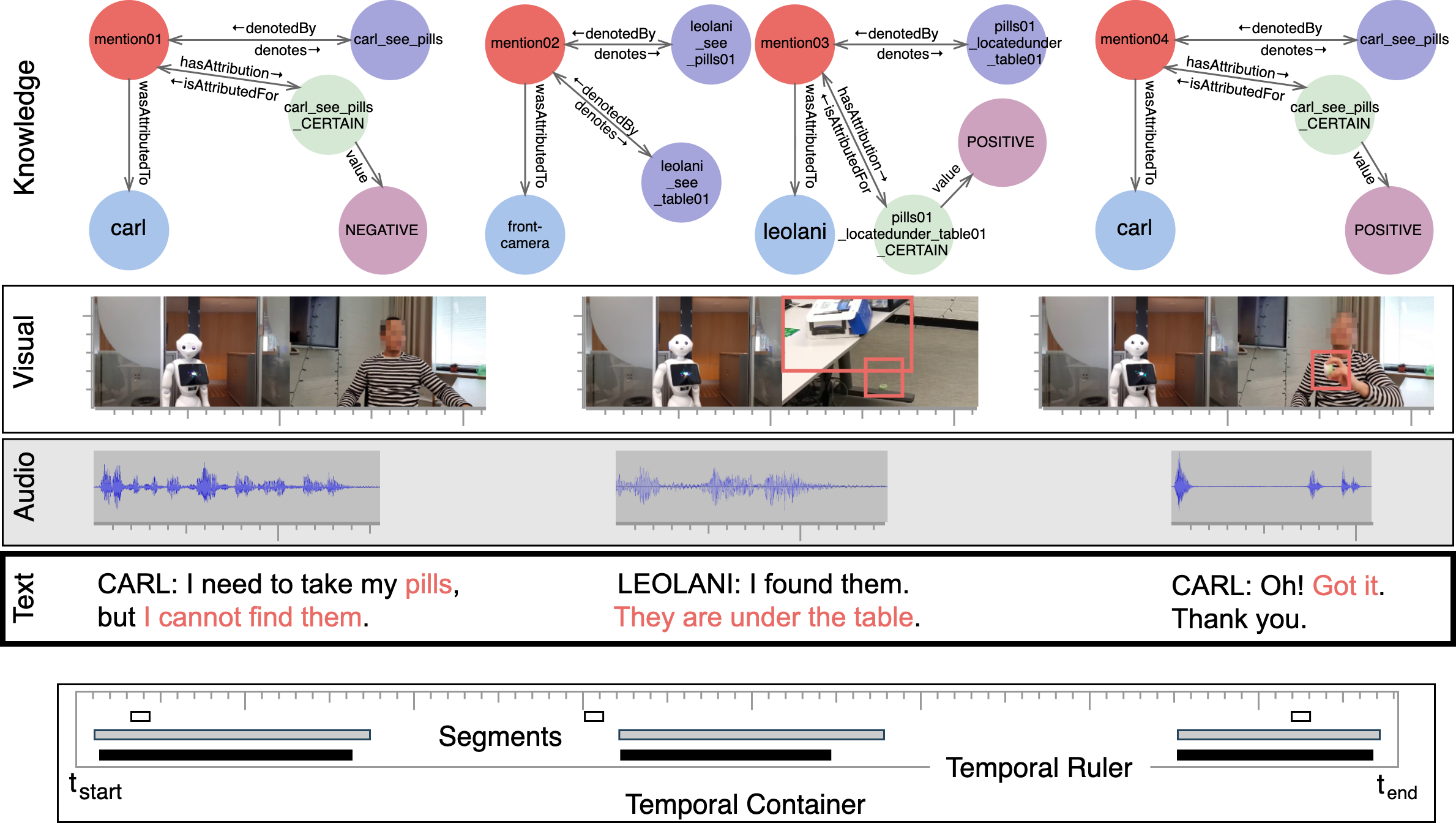}
  \caption{A visualization of the four modalities (text, audio, visual, and knowledge) from the CarLani example scenario. Signals are grounded in the temporal container shown on the horizontal axis, with bars marking their alignment through the temporal ruler. The red boxes mark \textit{segments} annotated as \textit{mentions} of objects (pills and the table). Text \textit{segments} highlighted in red have been annotated as \textit{mentions} of triples. The upper graphs represent corresponding triples from eKG automatically generated from the annotated source modalities along the temporal sequence. In the visual modality, two different camera viewpoints (left is what Carl sees and right is what Leolani sees) were concatenated side by side for the visualization.  %
  }
  \label{fig:swimming_pool}
\end{figure*}

\paragraph{Modalities}
The different modalities are represented in parallel streams of signals that are aligned by temporal and spatial rulers in the containers (Figure \ref{fig:swimming_pool}). We currently support text, audio, and visual modalities enriched by the knowledge layer as extracted by annotations.

\begin{example}
\begin{verbatim}
carl (scenariofolder)
    audio (folder)
    image (folder)
    video (folder)
    rdf (folder)
    text (folder)
    audio.json, image.json,
    video.json, text.json, carl.json
\end{verbatim}
\caption{Scenario file structure}
\label{ex2}
\end{example}

Within each modality, a signal is broken down into \textit{segments} positioned relative to the temporal and spatial ruler through begin and end points or box coordinates respectively. The granularity can vary but depends on the minimal unit of the rulers. Figure~\ref{fig:swimming_pool} shows an example of a scenario with layers for these four modalities with a temporal ruler on the horizontal axis. In this scenario, a person, "Carl", tells a robot, "Leolani", that he cannot find a pillbox. The robot spots the box under the table and communicates this to Carl, who confirms finding it. At every turn in the conversation, we see interaction data as segments (bars) aligned through the temporal ruler and its corresponding subgraphs generated from the interpretations as added to the eKG. The triples in the subgraph not only contain the representations for the participants and the pillbox but also representations of their mentions in the lower layers by specific sources. These mentions specify the offsets and box coordinates in the segments so that the graph can be related to the signal in time. The graphs also show that sources express denial and uncertainty through mentions. Our ToM model supports reasoning over the status of the triples derived from the signals: who said what, what sensor perceived what, etc.

The example furthermore shows that segments are not necessarily fully aligned but are always temporally ordered. At the second turn, the agent first perceives a pillbox under the table (white bar in the ruler), after that the agent reports this in audio and text (gray and black bar in the ruler).

It is possible to represent a scenario from recordings, as was done for the CarLani scenario in Figure~\ref{fig:swimming_pool}, but also to create these manually by simply adding series of audiovisual and/or text content as files to the scenario folder. However, every data representation needs to have a temporal ruler to ground all units in each modality to the same time period, which needs to be done in the corresponding JSON file for each modality. These JSON files can be created through scripts and the annotation tool described in Section~\ref{sec:annotation}. Optionally, there can be spatial rulers grounding the data in spatial areas.

\paragraph{Annotations and mentions}
Any segment can be annotated, which results in mentions or perceptions added to the JSON-LD file for the specific media. Mentions define a relation between text segments and interpretations, whereas perceptions relate audiovisual segments to interpretations. Each annotation has the following attributes: 1) type: kind of annotation; 2) value: the actual interpretation (e.g. label, reference); 3) source: software or person that created the annotation; 4) timestamp: when the annotation was created. We can have any number of segments with any number of annotations defined per mention/perception. Furthermore, annotations can be added on top of other annotations, following the Layered Annotation Framework \cite{ide2007towards}. Finally, annotations of segments in different modalities with the same identifier from the eKG will automatically create cross-modality co-reference; for example, in the CarLani scenario, the perception of the pillbox and its mention in  the utterances are mapped to the same instance URI in the eKG.

\paragraph{Identities, properties, and relations}
The core idea is to create a mapping between a segment of a \textit{signal} (e.g., a bounding box in an image or an offset position and length in a text) and the \textit{signal's interpretation} (e.g., a person's face or a pronoun).  %
Through referential grounding, we can generate triples expressing properties and relations across different modalities. Shared identifiers (URIs) aggregate these properties and relations, resulting in a world model over time. In Figure~\ref{fig:swimming_pool}, this is demonstrated by the sequence of subgraphs at the top showing different states of interpretation going from lack of knowledge about the location of the pillbox (negative polarity) to the perception and having it in possession (positive polarity). The triples stored in an eKG likewise reflect this accumulation over time, while each triple is still grounded to a segment in a modality. By faceting the triples in time, they also model emotional changes, such as smiling first and being angry next.

As explained in previous work \cite{vossen2018leolani, vossen_leolani_2019}, our framework focuses on storing information related to episodic experiences and their interpretations as perspectives that agents have. By nature, our framework is flexible in dealing with incomplete or contradicting information and can reason over knowledge with uncertainty while considering the sources' trustworthiness. For the scenario in Figure~\ref{fig:swimming_pool}, the model represents two realities at the same time point: ''Carl" not knowing the location and ''Leolani" knowing the location. Querying the model for the location of the pillbox at time t1 generates an answer according to ''Leolani". Reasoning is thus not only used to derive knowledge from knowledge or answer factual questions, but also to evaluate the quality of the knowledge itself. Agents can use such qualitative evaluations to formulate strategies and actions to improve knowledge states.

Following the principles of Linked Data, our framework reuses existing ontologies for provenance (PROV-O), text processing (NAF), event (SEM) and perspective modeling (GRaSP and GAF). Furthermore, the usage of RDF allows us to integrate information from other existing open Knowledge Graphs, for example, by querying WikiData or DBpedia to include prior knowledge (background or commonsense). Using JSON-LD elements in our data representation enables us to directly attach the referential grounding to the eKG by mapping elements of our JSON structure described in Figure \ref{fig:overview} to elements of the underlying ontologies of the eKG without losing the lightweight representation of plain JSON.

\section{Example Data Sets}
\label{sec:examples}

Any multimodal interaction data can be represented and annotated in the EMISSOR annotation format with minimal effort. We released EMISSOR representations and annotations of some popular public data sets (e.g. MELD and IEMOCAP), together with the scripts to convert them from their original formats. In the near future, we will add scripts for other popular data sets. In addition to the conversion scripts, we also created scripts for segmentation of modalities, such as bounding boxes for objects, faces and text tokenization with named entity detection. An additional baseline script resolves the identities of faces and entities against an eKG by selecting the first matching name. These scripts prepare any video recording or collection of multimedia data for annotation in the annotation tool described in Section~\ref{sec:annotation}. In the future, we replace the baseline scripts with SOTA modules for resolving referential ambiguity.

We also created our own data set called CarLani directly rendered from interacting with our robot platform. Figure~\ref{fig:swimming_pool} shows an example dialogue of three utterances from this data set between the human (Carl) and the robot (Leolani), assuming the context of a care taking robot in an elderly home.

By running interactions through our robot platform with humans in a physically perceived world, the multimodal data is automatically grounded in the knowledge graph according to the EMISSOR framework. This will automatically generate rich referential relations between mentions and perceptions with identities, within a functional communicative contexts. These can be analysed, evaluated and adapted to gold annotations for training and testing.

\section{Annotation Tool}
\label{sec:annotation}

Along with the proposed data representation, we are developing a GUI tool capable of reading EMISSOR data representations (with or without annotations).
The purpose of the tool is a first inspection of data sets by providing a comprehensible visualisation of the signals in different modalities, their grounding to the temporal (and spatial) containers, as well as their interpretations, including segment alignments, situated references and explicit semantic representations.
Second, it allows modification of the aforementioned properties for a given data set, e.g. to add gold annotations, perform corrections or add additional interpretations.
Third, gold scenarios for a given task or problem can be created manually from scratch without the need for an actual agent implementation.

Besides conversion issues to other data representations\footnote{
  Anvil supports DIAML and stores conversational units in sequences with name
  identifiers and time stamps for an associated video file. ELAN stores
  conversations in EAF (Eudico Annotation Format), which is a propriety XML
  format.
}, existing tools like e.g. Anvil\footnote{\url{www.anvil-software.org}} \cite{kipp2001anvil} or Elan\footnote{\url{www.mpi.nl/corpus/html/elan/}} \cite{brugman2004annotating} only ground the conversation to speakers, audio, faces, gestures but do not ground referential expressions to the situation. Our tool focuses on segmentation and grounding to mediate between the media data and the identities in the Knowledge Graph.

The current version of the tool supports image, audio, and text as modalities in a scenario, allows to add and remove signals to them and to position (ground) the signals on (to) the timeline of the scenario. In any situation, it is possible to create segments and annotations automatically or manually. On image signals, rectangular segments (bounding boxes) can be defined manually and annotated. Alternatively, during data preparation, boxing scripts can be used to automatically generate bounding boxes beforehand. Text signals can be automatically tokenized using scripts beforehand as well. The tool then allows for token selections to be annotated. The tool also provide choices for reference linking to known (listed) entities as annotation values. These entities can be taken from the eKG or from any other registration. In addition, you can create new identities directly in the tool through annotations, as well create triples that express properties or relations between entities. The URIs and triples created during the annotation can then be added as gold knowledge to the eKG a posteriori.

\begin{figure}
  \centering
  \includegraphics[width=\linewidth]{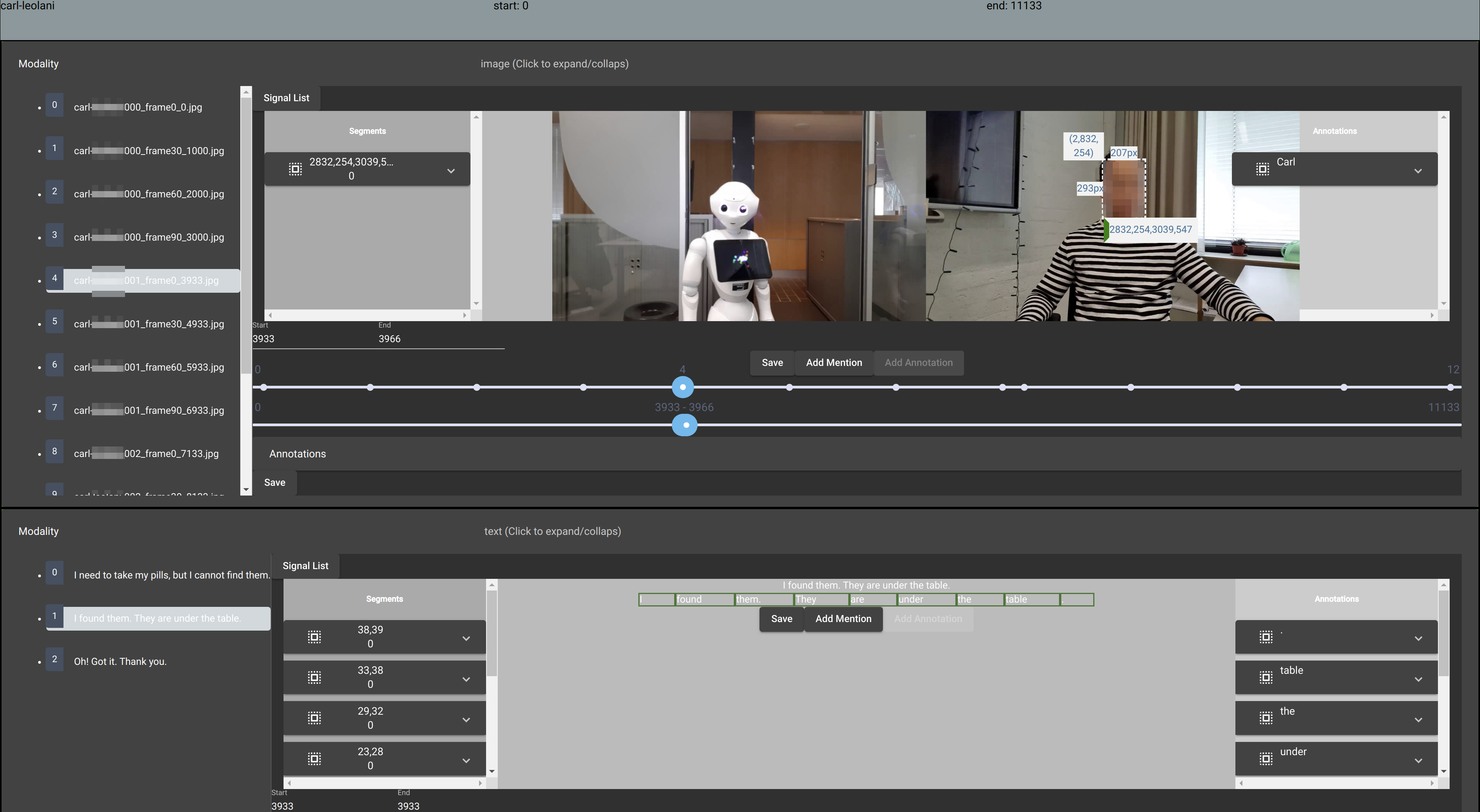}
  \caption{Tool for visualizing and annotating EMISSOR}
  \label{fig:tool_carl}
\end{figure}

Finally, the annotation tool can be used to create scenarios manually in a very controlled way. Researchers can store images and conversations in the corresponding media folders manually and next use the tool to place them in the proper order. In the near future, we will add a function to play such a scenario as well following the temporal specification. Currently, the user can play it by moving forward manually.

\section{Conclusions}
\label{sec:conclusion}

In this paper, we described eight desiderata for the representation of multimodal interactions to develop and test systems that can act intelligently in collaborative contexts. We argued that most representations do not satisfy all these desiderata. We therefore presented our EMISSOR platform for referential interpretation of multimodal interaction data to yield a Knowledge Graph that forms an explicit episodic memory of the experiences (eKG) which can be used for reasoning. EMISSOR combines light-weight JSON-LD representations for sequential media data with semantic web-based RDF models of an interpreted world. Through this we can model the cumulative growth of knowledge and information in the Knowledge Graph as a result of processing multimedia streams over time. EMISSOR is designed to address all eight desiderata. It will enable to create and compare recordings, annotations and interpretations of interactions in real-world contexts. This allows researchers to more easily share experiments and compare the interactions across different experiments and with different purposes, regardless of the specifics of the agents' systems or humans that participate in the experiment. Our model, software and converted data sets are available \footnote{\url{https://github.com/cltl/EMISSOR}} according to the Apache open source license.

As next steps, we are developing more tooling and the automatic linking of multimodal segments to identities. We will also provide more data by either converting existing public data to our framework or by rendering  data through our robot platform. So far we focused on grounding segments in temporal containers but not yet in spatial containers. Our interactions do not include motion and navigation. In future work, we hope to also include spatial grounding and reasoning. Finally, we want to include an evaluation framework for analysing system performance in relation to 1) qualitative properties of the interaction, 2) goals and intentions following a reinforcement learning approach and 3) by evaluating the resulting eKG.

\section{Acknowledgements}
This research was funded by the Vrije Universiteit Amsterdam and the Netherlands Organisation for Scientific Research (NWO) via the \textit{Spinoza} grant (SPI 63-260) awarded to Piek Vossen, the \textit{Hybrid Intelligence Centre} via the Zwaartekracht grant (024.004.022) and the research project (9406.D1.19.005) \textit{Communicating with and Relating to Social Robots: Alice Meets Leolani}.

\bibliography{anthology,acl2021,myref}
\bibliographystyle{acl_natbib}

\appendix
\onecolumn

\colorlet{punct}{red!60!black}
\definecolor{background}{HTML}{EEEEEE}
\definecolor{delim}{RGB}{20,105,176}
\colorlet{numb}{magenta!60!black}
\lstdefinelanguage{json}{
    basicstyle=\normalfont\ttfamily,
    numbers=left,
    numberstyle=\scriptsize,
    stepnumber=1,
    numbersep=7pt,
    showstringspaces=false,
    breaklines=true,
    frame=lines,
    backgroundcolor=\color{background},
    literate=
     *{0}{{{\color{numb}0}}}{1}
      {1}{{{\color{numb}1}}}{1}
      {2}{{{\color{numb}2}}}{1}
      {3}{{{\color{numb}3}}}{1}
      {4}{{{\color{numb}4}}}{1}
      {5}{{{\color{numb}5}}}{1}
      {6}{{{\color{numb}6}}}{1}
      {7}{{{\color{numb}7}}}{1}
      {8}{{{\color{numb}8}}}{1}
      {9}{{{\color{numb}9}}}{1}
      {:}{{{\color{punct}{:}}}}{1}
      {,}{{{\color{punct}{,}}}}{1}
      {\{}{{{\color{delim}{\{}}}}{1}
      {\}}{{{\color{delim}{\}}}}}{1}
      {[}{{{\color{delim}{[}}}}{1}
      {]}{{{\color{delim}{]}}}}{1},
}

\section*{Appendix A: CarlLani Example Data from redacted source}

In the following we include an excerpt of the CarlLani data set as the original source was redacted from the submission for anonymization purposes. For space and readability reasons {\em text.json} and {\em image.json} metadata files are shortened by removing part of the signals and/or mentions.
Also the context referenced in the JSON-LD {\em  @context} element is included.

\subsection*{Scenario structure}
\begin{Verbatim}[baselinestretch=0.1]
|- carl-robot/
  |- image/
    |- carl-robot-000_frame0_0.jpg
    |- carl-robot-000_frame30_1000.jpg
    |- carl-robot-000_frame60_2000.jpg
    |- ....
  |- rdf/
    |- statement1.trig
    |- objectdetection1.trig
    |- statement2.trig
    |- statement3.trig
  |- text/
    |- carl-robot.csv
  |- carl-robot.json
  |- image.json
  |- text.json
\end{Verbatim}

\subsection*{carl-robot.csv}
\begin{Verbatim}[fontsize=\small,baselinestretch=0.1]
speaker,utterance,time
Carl,"I need to take my pills, but I cannot find them.",0
Leolani,"I found them. They are under the table.",3933
Carl,"Oh! Got it. Thank you.",7133
\end{Verbatim}

\subsection*{carl-robot.json}
\begin{lstlisting}[language=json,firstnumber=1]
{
  "@context" : "http://emissor.org/jsonldcontext.jsonld",
  "type": "Scenario",
  "id": "carl-robot",
  "context": {
    "agent": "robot_agent",
    "objects": [],
    "persons": [],
    "speaker": {
      "@context" : "http://schema.org/docs/jsonldcontext.jsonld",
      "id": "bc913d64-a597-4876-a3fe-fe47472cd274",
      "type": "Person",
      "birthDate": "1995-04-09T20:00:00Z",
      "gender": "Male",
      "name": "Carl"
    }
  },
  "ruler": {
    "type": "TemporalRuler",
    "container_id": "carl-robot",
    "end": 11133,
    "start": 0
  },
  "signals": {
    "image": "./image.json",
    "text": "./text.json"
  }
}
\end{lstlisting}

\newpage
\subsection*{image.json (excerpt)}
\begin{lstlisting}[language=json,firstnumber=1]
[{
  "@context" : "http://emissor.org/jsonldcontext.jsonld",
  "type": "ImageSignal",
  "id": "21830691-4410-45f2-b611-f61cb4dbc0de",
  "files": [
    "image/carl-robot-000_frame0_0.jpg"
  ],
  "modality": "image",
  "time": {
    "type": "TimeSegment",
    "container_id": "carl-robot",
    "start": 0,
    "end": 33
  },
  "ruler": {
    "type": "MultiIndex",
    "container_id": "21830691-4410-45f2-b611-f61cb4dbc0de",
    "bounds": [0, 0, 3840, 1080]
  },
  "mentions": [
    {
      "type": "Mention",
      "id": "54920da9-41d4-421e-b3f4-7955e71f053a",
      "annotations": [
        {
          "type": "Annotation",
          "source": "machine",
          "timestamp": 0,
          "type": "person",
          "value": {
            "type": "Face",
            "instance": {
              "@context" : "http://schema.org/docs/jsonldcontext.jsonld",
              "id": "bc913d64-a597-4876-a3fe-fe47472cd274",
              "type": "Person",
              "birthDate": "1995-04-09T20:00:00Z",
              "gender": "Male",
              "name": "Speaker"
            },
            "age": 23,
            "gender": "male",
            "faceprob": 1.0
          }
        }
      ],
      "segment": [
        {
          "type": "BoundingBox",
          "container_id": "21830691-4410-45f2-b611-f61cb4dbc0de",
          "bounds": [2830, 241, 3034, 521]
        }
      ]
    }
  ]},





  {
    "@context" : "http://emissor.org/jsonldcontext.jsonld",
    "type": "ImageSignal",
    "id": "88a31791-4410-45f2-b611-f61cb4d321ff",
    "files": [
      "image/carl-robot-000_frame30_1000.jpg"
    ],
    "modality": "image",
    "time": {
      "type": "TimeSegment",
      "container_id": "carl-robot",
      "start": 1000,
      "end": 1033
    },
    "ruler": {
      "type": "MultiIndex",
      "container_id": "88a31791-4410-45f2-b611-f61cb4d321ff",
      "bounds": [0, 0, 3840, 1080]
    },
    "mentions": [
      {
        "type": "Mention",
        "id": "92af1ea9-41d4-421e-b3f4-7955e71a1a97",
        "annotations": [
          {
            "type": "Annotation",
            "source": "machine",
            "timestamp": 1000,
            "type": "person",
            "value": {
              "type": "Face",
              "instance": {
                "@context" : "http://schema.org/docs/jsonldcontext.jsonld",
                "@id": "bc913d64-a597-4876-a3fe-fe47472cd274",
                "type": "Person",
                "birthDate": "1995-04-09T20:00:00Z",
                "gender": "Male",
                "name": "Speaker"
              },
              "age": 21,
              "gender": "male",
              "faceprob": 1.0
            }
          }
        ],
        "segment": [
          {
            "type": "BoundingBox",
            "container_id": "88a31791-4410-45f2-b611-f61cb4d321ff",
            "bounds": [2831, 235, 3036, 514]
          }
        ]
      }]}, .....]
\end{lstlisting}

\newpage
\subsection*{text.json (excerpt)}
\begin{lstlisting}[language=json,firstnumber=1,basicstyle=\scriptsize]
[{
    "@context" : "http://emissor.org/jsonldcontext.jsonld",
    "files": ["text/carl-robot.csv#0"],
    "id": "85c27957-9b18-497e-9557-761b02bdbc21",
    "mentions": [
      {
        "type": "Mention",
        "id": "0d830564-ab25-4aac-82f6-f34fc61b0481",
        "annotations": [
          {
            "source": "annotation_tool",
            "timestamp": 1616442473,
            "type": "token",
            "value": {
              "id": "b1ec4a11-cd35-4c10-be47-244147da1086",
              "ruler": {
                "container_id": "b1ec4a11-cd35-4c10-be47-244147da1086",
                "type": "AtomicRuler"
              },
              "type": "Token",
              "value": "I"
            }
          }
        ],
        "segment": [
          {
            "container_id": "85c27957-9b18-497e-9557-761b02bdbc21",
            "start": 0,
            "stop": 1,
            "type": "Index"
          }
        ]
      },
      ....
      {
        "type": "Mention",
        "id": "a930c234-f3f2-4932-a32d-bde0acc2aafd",
        "annotations": [
          {
            "source": "annotation_tool",
            "timestamp": 1616442473,
            "type": "token",
            "value": {
              "id": "13d77c30-4f10-481a-b0c4-3b80532b038f",
              "ruler": {
                "container_id": "13d77c30-4f10-481a-b0c4-3b80532b038f",
                "type": "AtomicRuler"
              },
              "type": "Token",
              "value": "."
            }
          }
        ],
        "segment": [
          {
            "container_id": "85c27957-9b18-497e-9557-761b02bdbc21",
            "start": 47,
            "stop": 48,
            "type": "Index"
          }
        ]
      }
    ],
    "modality": "text",
    "ruler": {
      "container_id": "85c27957-9b18-497e-9557-761b02bdbc21",
      "start": 0,
      "stop": 48,
      "type": "Index"
    },
    "seq":["I"," ","n","e","e","d"," ","t","o"," ","t","a","k","e"," ","m","y"," ","p","i","l","l","s",",",
    " ","b","u","t"," ","I"," ","c","a","n","n","o","t"," ","f","i","n","d"," ","t","h","e","m","."],
    "time": {
      "container_id": "carl-robot",
      "end": 0,
      "start": 0,
      "type": "TemporalRuler"
    },
    "type": "TextSignal"
  },
  ....







  ....
  {
    "@context" : "http://emissor.org/jsonldcontext.jsonld",
    "files": [
      "text/carl-robot.csv#2"
    ],
    "id": "2142b6d8-4cda-481b-a056-1b6d874da648",
    "mentions": [
      {
        "type": "Mention",
        "id": "c851ca48-81b6-44fe-a772-9f62840ca2f6",
        "annotations": [
          {
            "source": "annotation_tool",
            "timestamp": 1616442473,
            "type": "token",
            "value": {
              "id": "d7770947-0be5-413f-9c1e-4e9d130e6a41",
              "ruler": {
                "container_id": "d7770947-0be5-413f-9c1e-4e9d130e6a41",
                "type": "AtomicRuler"
              },
              "type": "Token",
              "value": "Oh"
            }
          }
        ],
        "segment": [
          {
            "container_id": "2142b6d8-4cda-481b-a056-1b6d874da648",
            "start": 0,
            "stop": 2,
            "type": "Index"
          }
        ]
      },
      ....
      {
        "type": "Mention",
        "id": "e62ae54b-bbb4-4464-8796-fe1a5ce22fac",
        "annotations": [
          {
            "source": "annotation_tool",
            "timestamp": 1616442473,
            "type": "token",
            "value": {
              "id": "fb7a3f36-11c4-486c-bd60-aeedd4377bb7",
              "ruler": {
                "container_id": "fb7a3f36-11c4-486c-bd60-aeedd4377bb7",
                "type": "AtomicRuler"
              },
              "type": "Token",
              "value": "."
            }
          }
        ],
        "segment": [
          {
            "container_id": "2142b6d8-4cda-481b-a056-1b6d874da648",
            "start": 21,
            "stop": 22,
            "type": "Index"
          }
        ]
      }
    ],
    "modality": "text",
    "ruler": {
      "container_id": "2142b6d8-4cda-481b-a056-1b6d874da648",
      "start": 0,
      "stop": 22,
      "type": "Index"
    },
    "seq": ["O","h","!"," ","G","o","t"," ","i","t","."," ","T","h","a","n","k"," ","y","o","u","."],
    "time": {
      "container_id": "carl-robot",
      "end": 7133,
      "start": 10976,
      "type": "TemporalRuler"
    },
    "type": "TextSignal"
  }
]

\end{lstlisting}

\newpage
\subsection*{JSON-LD context (http://emissor.org/jsonldcontext.jsonld)}
\begin{lstlisting}[language=json,firstnumber=1]
{
  "@context" : {
    "@base": "http://experiment.my/",
    "@vocab": "https://emmisor.org/emissor#",
    "type": "@type",
    "id": "@id",
    "emissor": "http://emmisor.org/emissor#",
    "grasp": "http://groundedannotationframework.org/grasp#",
    "container_id": {"@type": "@id"},
    "signal": "@nest",
    "Mention": "grasp:Mention"
  }
}
\end{lstlisting}

\newpage

\section*{statements2.trig}

\begin{lstlisting}[language=json,firstnumber=1,basicstyle=\scriptsize]
@prefix robotContext: <http://emissor.org/robot/context/> .
@prefix xml1: <https://www.w3.org/TR/xmlschema-2/#> .
@prefix owl: <http://www.w3.org/2002/07/owl#> .
@prefix wdt: <http://www.wikidata.org/prop/direct/> .
@prefix ceo: <http://www.newsreader-project.eu/domain-ontology#> .
@prefix gaf: <http://groundedannotationframework.org/gaf#> .
@prefix ns1: <urn:x-rdflib:> .
@prefix wd: <http://www.wikidata.org/entity/> .
@prefix grasp: <http://groundedannotationframework.org/grasp#> .
@prefix xml: <http://www.w3.org/XML/1998/namespace> .
@prefix grasps: <http://groundedannotationframework.org/grasp/sentiment#> .
@prefix sem: <http://semanticweb.cs.vu.nl/2009/11/sem/> .
@prefix prov: <http://www.w3.org/ns/prov#> .
....
@prefix foaf: <http://xmlns.com/foaf/0.1/> .
@prefix wgs: <http://www.w3.org/2003/01/geo/wgs84_pos#> .
@prefix graspf: <http://groundedannotationframework.org/grasp/factuality#> .
@prefix xsd: <http://www.w3.org/2001/XMLSchema#> .
@prefix rdfs: <http://www.w3.org/2000/01/rdf-schema#> .
@prefix grasp: <http://groundedannotationframework.org/grasp#> .


robotWorld:Instances {
  robotWorld:lani a gaf:Instance, robotMu:robot ;
                  rdfs:label "lani" .
  robotWorld:pills a gaf:Instance, robotMu:object ;
                  rdfs:label "pills" ;
                  gaf:denotedIn robotTalk:chat1_utterance2_char0-39 .
  robotWorld:pills-277239 a gaf:Instance, robotMu:object, robotMu:pills ;
                  rdfs:label "pills-277239" ;
                  robotMu:id "277239"^^xml1:string ;
                  gaf:denotedIn robotTalk:visual1_detection2_pixel0-3  ;
                  eps:hasContext robotContext:context212127 .
  robotWorld:table a gaf:Instance, robotMu:object ;
                  rdfs:label "table" ;
                  gaf:denotedIn robotTalk:chat1_utterance2_char0-39 .
  robotWorld:table-208510 a gaf:Instance, robotMu:object, robotMu:table ;
                  rdfs:label "table-208510" ;
                  robotMu:id "208510"^^xml1:string ;
                  gaf:denotedIn robotTalk:visual1_detection2_pixel0-3 ;
                  eps:hasContext robotContext:context212127 .
}

robotTalk:Interactions {
  robotWorld:Netherlands a robotMu:location, sem:Place, robotMu:country ;
                  rdfs:label "Netherlands" .
  robotWorld:Gelderland a robotMu:location, sem:Place, robotMu:region ;
                  rdfs:label "Gelderland" .
  robotWorld:Apeldoorn a robotMu:location, sem:Place, robotMu:city ;
                  rdfs:label "Apeldoorn" .
  robotTalk:chat1 a sem:Event, grasp:Chat  ;
                  rdfs:label "chat1"  ;
                  robotMu:id "1"^^xml1:string  ;
                  sem:hasSubEvent robotTalk:chat1_utterance2 .
  robotTalk:visual1 a sem:Event, grasp:Visual ;
                  rdfs:label "visual1" ;
                  robotMu:id "1"^^xml1:string ;
                  sem:hasSubEvent robotTalk:visual1_detection2 .
  robotTalk:chat1_utterance2 a sem:Event, grasp:Utterance ;
                  rdfs:label "chat1_utterance2" ;
                  robotMu:id "2"^^xml1:string ;
                  sem:hasActor robotFriends:lani .
  robotTalk:visual1_detection2 a sem:Event, grasp:Detection ;
                  rdfs:label "visual1_detection2" .
                  robotMu:id "2"^^xml1:string ;
                  sem:hasActor robotInputs:front-camera .
  robotInputs:front-camera a gaf:Instance, grasp:Source, sem:Actor, robotMu:sensor ;
                  rdfs:label "front-camera" .
  robotFriends:lani a robotMu:person, gaf:Instance, grasp:Source, sem:Actor ;
                  rdfs:label "lani" .
  robotContext:home a robotMu:location, sem:Place ;
                  rdfs:label "home" ;
                  robotMu:id "251375"^^xml1:string ;
                  robotMu:in robotWorld:Netherlands, robotWorld:Gelderland, robotWorld:Apeldoorn .
  robotContext:context212127 a eps:Context ;
                  rdfs:label "context212127" ;
                  robotMu:id "212127"^^xml1:string ;
                  eps:hasDetection robotWorld:pills-277239, robotWorld:table-208510 ;
                  sem:hasBeginTimeStamp robotContext:2021-03-12;
                  sem:hasEvent robotTalk:chat1, robotTalk:visual1;
                  sem:hasPlace robotContext:home .
  robotContext:2021-03-12 a sem:Time, time:DateTimeDescription ;
                  rdfs:label "2021-03-12" ;
                  time:day "12"^^xml1:gDay ;
                  time:month "3"^^xml1:gMonthDay ;
                  time:unitType time:unitDay ;
                  time:year "2021"^^xml1:gYear .
}



robotWorld:Claims {
  robotWorld:lani_sense_front-camera a gaf:Assertion, sem:Event ;
                  rdfs:label "lani_sense_front-camera" .
  robotWorld:lani_know_lani a gaf:Assertion, sem:Event ;
                  rdfs:label "lani_know_lani" ;
                  owl:sameAs robotWorld:lani .
  robotWorld:pills_locatedunder_table a gaf:Assertion, sem:Event ;
                  rdfs:label "pills_locatedunder_table" ;
                  gaf:denotedBy robotTalk:chat1_utterance2_char0-39 .
  robotWorld:lani_see_pills-277239 a gaf:Assertion, sem:Event ;
                  rdfs:label "lani_see_pills-277239" ;
                  gaf:denotedBy robotTalk:visual1_detection2_pixel0-3 ;
                  eps:hasContext robotContext:context212127 .
  robotWorld:lani_see_table-208510 a gaf:Assertion, sem:Event ;
                  rdfs:label "lani_see_table-208510" ;
                  gaf:denotedBy robotTalk:visual1_detection2_pixel0-3 ;
                  eps:hasContext robotContext:context212127 .
}

robotTalk:Perspectives {
  robotTalk:chat1_utterance2_char0-39 a gaf:Mention, grasp:Statement ;
                  rdfs:label "chat1_utterance2_char0-39"
                  rdf:value "I found them. They are under the table."^^xml1:string .
                  prov:wasDerivedFrom robotTalk:chat1_utterance2 ;
                  gaf:denotes robotWorld:pills_locatedunder_table ;
                  gaf:containsDenotation robotWorld:pills, robotWorld:table ;
                  grasp:wasAttributedTo robotFriends:lani ;
                  grasp:hasAttribution robotTalk:pills_locatedunder_table_CERTAIN-POSITIVE-NEUTRAL-NEUTRAL .
  robotTalk:visual1_detection2_pixel0-3 a gaf:Mention, grasp:Experience ;
                  rdfs:label "visual1_detection2_pixel0-3" ;
                  prov:wasDerivedFrom robotTalk:visual1_detection2 .
                  gaf:denotes robotWorld:lani_see_pills-277239, robotWorld:lani_see_table-208510 ;
                  gaf:containsDenotation robotWorld:pills-277239, robotWorld:table-208510 ;
                  grasp:wasAttributedTo robotInputs:front-camera ;
                  grasp:hasAttribution robotTalk:pills_locatedunder_table_PROBABLE .
  robotTalk:pills_locatedunder_table_CERTAIN-POSITIVE-NEUTRAL-NEUTRAL a grasp:Attribution ;
                  rdfs:label "pills_locatedunder_table_CERTAIN-POSITIVE-NEUTRAL-NEUTRAL" ;
                  rdf:value graspf:CERTAIN, graspf:POSITIVE, graspe:NEUTRAL, grasps:NEUTRAL ;
                  grasp:isAttributionFor robotTalk:chat1_utterance2_char0-39 .
  robotTalk:pills_locatedunder_table_PROBABLE a grasp:Attribution ;
                  rdfs:label "pills_locatedunder_table_PROBABLE" ;
                  rdf:value graspf:PROBABLE ;
                  grasp:isAttributionFor robotTalk:visual1_detection2_pixel0-3 .
  graspe:NEUTRAL a grasp:AttributionValue, graspe:EmotionValue .
  grasps:NEUTRAL a grasp:AttributionValue, grasps:SentimentValue .
  graspf:CERTAIN a grasp:AttributionValue, graspf:CertaintyValue .
  graspf:POSITIVE a grasp:AttributionValue, graspf:PolarityValue .
  graspf:PROBABLE a grasp:AttributionValue, graspf:CertaintyValue .
}

robotWorld:lani_know_lani {
  robotWorld:lani robotMu:know robotFriends:lani .
}

robotWorld:lani_sense_front-camera {
  robotWorld:lani robotMu:sense robotInputs:front-camera .
}

robotWorld:pills_locatedunder_table {
  robotWorld:pills robotMu:locatedUnder robotWorld:table .
}

robotWorld:lani_see_pills-277239 {
  robotWorld:lani robotMu:see robotWorld:pills-277239 .
}

robotWorld:lani_see_table-208510 {
  robotWorld:lani robotMu:see robotWorld:table-208510 .
}
\end{lstlisting}

\end{document}